\documentstyle[aps]{revtex}
\begin{document}
\title{
Determining the Saddle Point in Micromagnetic Models of Magnetization Switching
}
\author{
Gregory Brown$^{1,2}$
and M.A.~Novotny$^{3}$
and Per Arne Rikvold$^{2,4}$
}
\address{
$^1$ Center for Computational Sciences, Oak Ridge National Laboratory,\\
P.O.Box 2008 Mail Stop 6114, Oak Ridge, TN 37831-6114, USA\\
$^2$ School of Computational Science and Information Technology,\\
Florida State University, Tallahassee, FL 32306-4120, USA\\
$^3$ P.O. Box 5167, Department of Physics and Astronomy, Mississippi State University,\\
Mississippi State, MS 39762-5167, USA\\
$^4$ Center for Materials Research and Technology and Department of
Physics,\\
Florida State University, Tallahassee, FL 32306-4351, USA\\
}

\maketitle    

\begin{abstract}
A numerical model of single-domain nanoscale iron magnets fabricated
using scanning-microscope-assisted chemical vapor deposition is
simulated using finite-temperature micromagnetics. A
Projective-dynamics method is used to determine the magnetization at
the saddle point as a function of temperature. This magnetization is
found to decrease linearly as the temperature is raised.
\end{abstract}

\vskip 0.1in

An interesting problem in nonequilibrium statistical mechanics, with
numerous applications in condensed-matter physics and materials
science, is how a system approaches the global minimum of the free
energy.  One simple, but common, situation is that of a local
free-energy minimum which is separated from the global minimum by a
free-energy barrier. In other words, any path between the metastable
local minimum and the equilibrium global minimum involves an initial
increase in the free energy. The most probable path between the minima
involves the minimum free energy increase since spontaneous increases
in free energy are improbable.  The maximum free energy along that most
probable path corresponds to the saddle point.  Often, the behavior of
the nonequilibrium dynamics are dominated by properties near the minima
and the saddle point. While the minima can be determined by examining
histograms of the state of the system, the saddle point is much harder
to determine. In this paper we present a technologically important
example of finding the saddle point using a projective-dynamics
technique.

The results described here are for a numerical model of single-domain
nanoscale iron magnets that have been fabricated using chemical vapor
deposition directed by a scanning-tunneling microscope \cite{WIRTH}.
These magnetic pillars have cross-sectional dimensions on the order of
$10$ nm and extend on the order of $100$ nm perpendicular
to the substrate.  The magnetic pillars are modeled by a
one-dimensional array of magnetization density vectors 
${{\vec M}}({{\vec r}})$ with fixed length $M_s$. Each vector precesses
around a local field ${{\vec H}}({{\vec r}})$ according to the 
Landau-Lifshitz-Gilbert equation \cite{BROWN63,AHARONI}
\begin{equation}
\frac{ {d} {{\vec M}}({{\vec r}}) }
     { {d} {t} }
 =
   \frac{ \gamma_0 }
        { 1+\alpha^2 }
   {{\vec M}}({{\vec r}})
 \times
 \left[
   {{\vec H}}({{\vec r}})
  -\frac{\alpha}{M_s} {{\vec M}}({{\vec r}}) \times 
                      {{\vec H}}({{\vec r}})
 \right]
\;,
\end{equation}
where the electron gyromagnetic ratio $\gamma_0 = 1.76\times10^7$
Hz/Oe. The phenomenological damping parameter $\alpha=0.1$ was
chosen to give underdamped dynamics. Properties of bulk iron were
assumed with the saturation magnetization $M_s=1700$  emu/cm$^3$
and the exchange length $\ell_x=2.6$ nm. The model pillar has a
square cross section with area $4 \ell_x^2,$ and is $34 \ell_x$ long.
Details of the numerical approach, including the inclusion of thermal
fluctuations in the local field, are given in Ref.~\cite{MMAG}.

These nanopillars have a strong shape anisotropy, and the
magnetization is most favorably oriented along the long axis of the
magnet, which is taken to be the $z$ direction.  The nanomagnets
are prepared in a metastable state via equilibration in an
externally-applied magnetic field $+H_0\hat{z}$, which is subsequently
varied rapidly in magnitude (without changing its orientation) to
$-H_0\hat{z}.$ If the field $H_0$ is less than the coercive field,
here $H_c \sim 1500$ Oe \cite{MMAG}, a free-energy barrier 
separates the positively-oriented magnetization from the equilibrium
negative orientation. Previous simulations \cite{MMAG} suggest that 
the saddle point in these pillars corresponds to the nucleation of 
a region of reversed magnetization at one of the ends of the pillar.
The nucleated region grows until the entire magnetization is reversed.

A useful technique for determining the magnetization corresponding to
the free-energy maximum of the barrier, also called the saddle point,
is the projective-dynamics technique \cite{KOLESIK,NOVOTNY,MITCHELL}.
The essence of this technique is projecting the original dynamics
described in a high-dimensional phase space onto a probabilistic
dynamic in a phase space of much lower dimension. A specific example
is helpful. The dynamics of an Ising model can be projected onto the
number of overturned spins, a measure of the global magnetization. The
dynamics of the one-dimensional model are then described in terms of
the probability that the number of overturned spins increases or
decreases. Thinking in terms of the number of spins in the stable
orientation, these probabilities are referred to as the probability of
growing ($P_{grow}$) and shrinking ($P_{shrink}$), respectively
\cite{KOLESIK}.

Projective dynamics for the model nanopillars is more complicated than
that for the Ising model for two reasons. First, the magnetization is
a continuous, three-dimensional variable. This is handled by
projecting the $z$-component of the global magnetization into
uniformly sized bins \cite{MITCHELL}. Then $P_{grow}$, the
probability of the region of stable magnetization ``growing,'' is the
probability that during an integration step the magnetization moves
into a bin corresponding to a smaller magnetization. Similarly,
$P_{shrink}$ is the probability that the magnetization moves into a
bin corresponding to a larger magnetization. The second complication
involves the persistent, subcritical regions of reduced magnetization
that develop at each end of the pillar. To accomodate this, during the
projective-dynamics analysis the pillar is divided into its top and bottom
parts with the normalized, ``global'' magnetization $M_z$ of each
half considered separately.

The measured $P_{grow}$ and $P_{shrink}$ at
$H_0$$=$$1000$ Oe are shown in Fig.~1 for temperatures of
$20$ K, $50$ K, and $100$ K. For each
temperature the results represent averaging over more than $10^9$
integration steps and on the order of $10^4$ switches.  Consider the
probabilistic dynamics at $M_z$$=$$0.85$ for $T$$=$$100$
K. Here $P_{shrink}$$>$$P_{grow}$ and on average the
magnetization moves to the right. For $M_z$$=$$0.98$ the situation is
reversed, and on average the magnetization moves to the left. The
crossing point where $P_{shrink}$$=$$P_{grow}$ corresponds to
a locally stable fixed point since, on average, here the
magnetization moves towards the crossing.  In this case the crossing
is the metastable local free-energy minimum. For the crossing near
$M_z$$=$$0.74$ the magnetization moves away from the point on average,
and it corresponds to an unstable local maximum in the free
energy. This is the saddle point. Measuring the value of $M_z$ for
which $P_{shrink}$$=$$P_{grow}$ is a convenient method for
finding the saddle point.

From Fig.~1 it is obvious that $M_z$ at the saddle point is
different for magnetization reversal at different temperatures. The
value of $M_z$ at the saddle point is presented vs temperature in
Fig.~2. The point where $P_{shrink}$$=$$P_{grow}$ is estimated
using lines fit via a least-squares method to each probability in
the region near the crossing. The error bars are taken large
enough to include nearly all $M_z$ for which the difference in the
probabilities is less than the fluctuations in their estimates.

There is a clear linear trend in $M_z$ at the saddle point with
respect to the temperature. The solid line in Fig.~2 is an unweighted
least-squares fit to the data. This fit indicates that $M_z(T
\rightarrow 0 )$$=$$0.860$ at the saddle point. The
dependence on temperature is quite strong, since $M_z(100$ K$)$$=$$0.72$.

In summary, micromagnetic simulations at finite temperature have been
used to investigate magnetization reversal in single-domain nanoscale
magnets.  The saddle point in the free energy was determined using the
projective-dynamics technique, and the magnetization at the saddle
point was found to depend linearly on the temperature for all the
temperatures investigated, $T$$\le$$100$ K.

This work was supported in part by the Laboratory Directed Research
and Development Program of Oak Ridge National Laboratory, managed by
UT-Battelle, LLC for the U.S. Department of Energy under Contract
No. DE-AC05-00OR22725, and by the Ames Laboratory, which is operated
for the U.S. Department of Energy by Iowa State University under
Contract No. W-7405-82. This work was also supported by NSF grant No.\
DMR-0120310 and by FSU/CSIT. Computer resources were provided by the 
FSU Deptartment of Physics.

\newpage
~
\begin{figure}[t]
\vspace{6.0truecm}
\includegraphics{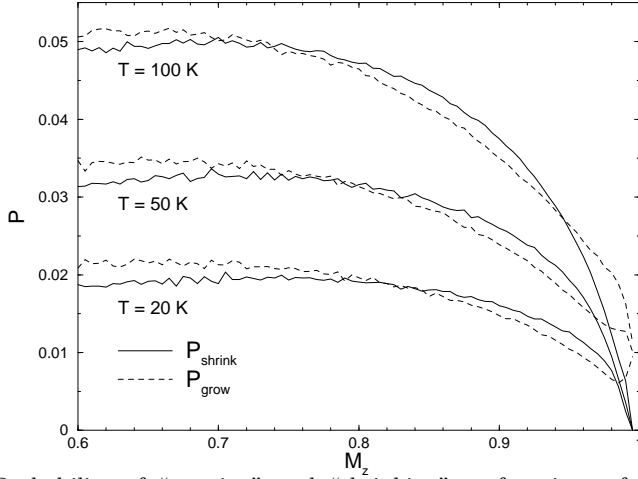}

\caption[] {Probability of ``growing'' and ``shrinking'' as
functions of the magnetization of one half of the pillar for three
temperatures: $100$ K, $50$ K, and $20$ K. The 
left-most crossing shown for a given temperature indicates the saddle
point, while the right-most crossing indicates the metastable 
local free-energy minimum. The crossings associated with the equilibrium
global free-energy minimum lie near $-1$ and are not shown}
\end{figure}

~
\begin{figure}[t]
\vspace{6.0truecm}
\includegraphics{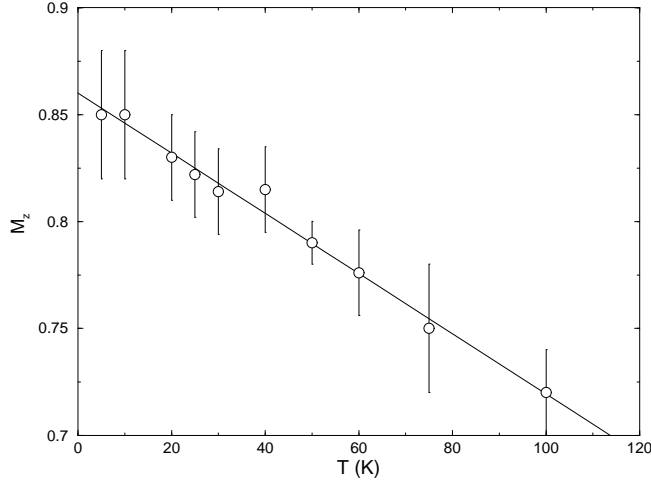}

\caption[] {Magnetization along the pillar, $M_z$ at the saddle point as
a function of temperature (circle). The magnetization at the saddle point 
decreases with increasing temperature, and the solid line is a least-squares
fit with intercept at $0.860$. The error bars are estimated from the range of
$M_z$ where $P_{grow}$ andn $P_{shrink}$ are nearly equal}
\end{figure}

\end{document}